\documentclass[conference]{IEEEtran}
\usepackage{graphicx}
\usepackage{amsmath}
\usepackage{cite}

\begin{document}

\title{\LARGE Automated identification of metamorphic test scenarios for an ocean-modeling application}

\author{
  \authorblockN{
    Dilip J. Hiremath\authorrefmark{1}\authorrefmark{2},
    Martin Claus\authorrefmark{1}\authorrefmark{2},
    Wilhelm Hasselbring\authorrefmark{2} and
    Willi Rath\authorrefmark{1}
  }
  \authorblockA{\authorrefmark{1}GEOMAR, Helmholtz Centre for Ocean Research, Kiel, Germany \\}
  \authorblockA{\authorrefmark{2}Kiel University, Software Engineering Group, Kiel, Germany\\}
}

\maketitle

\begin{abstract}

Metamorphic testing seeks to validate software in the absence of test oracles.
Our application domain is ocean modeling, where test oracles often do not exist, but where symmetries of the simulated physical systems are known.
In this short paper we present work in progress for automated generation of metamorphic test scenarios using machine learning.

Metamorphic testing may be expressed as f(g(X))=h(f(X)) with f being the application under test, with input data X, and with the metamorphic relation (g, h).
Automatically generated metamorphic relations can be used for constructing regression tests, and for comparing different versions of the same software application.

Here, we restrict to h being the identity map. Then, the task of constructing tests means finding different g which we tackle using machine learning algorithms.
These algorithms typically minimize a cost function.
As one possible g is already known to be the identity map, for finding a second possible g, we construct the cost function to minimize for g being a metamorphic relation and to penalize for g being the identity map.
After identifying the first metamorphic relation, the procedure is repeated with a cost function rewarding g that are orthogonal to previously found metamorphic relations.

For experimental evaluation, two implementations of an ocean-modeling application will be subjected to the proposed method with the objective of presenting the use of metamorphic relations to test the implementations of the applications.

\end{abstract}

\begin{keywords}
Metamorphic testing, Ocean-modeling application testing, Oracle problem, Metamorphic relation, Test case generation, Software testing
\end{keywords}

\section{Motivation}

Ocean-modeling applications often build on legacy codes that usually do not feature any systematic approach to software testing~\cite{Johanson2018}.
Establishing tests for such complex systems is a challenge that is complicated further by the fact that the exploratory nature of scientific software often makes it impossible to have test oracles for specific results.
Alternatively, symmetries of the physical system that is modeled can be used to formulate necessary conditions to check for the correctness of the software. These correctness measure will not be based on results, but on the software's behaviour under changes to the input data.

Metamorphic testing is an approach to the generation of test cases and the verification of test results.
The central element of Metamorphic testing is a set of metamorphic relations. 
Metamorphic relations are necessary properties of the target application in relation to multiple inputs and their corresponding outputs \cite{Chen2018}. 
We intend to use machine learning for identifying metamorphic relations and derive metamorphic tests to validate the ocean system models \cite{Kanewala2013}.
Here, we use a simple but realistic Ocean-modelling application to demonstrate our approach.

Once identified, metamorphic relations can not only be used to construct regression tests for software that is under development but also for validating different implementations of the same software.
Further, since the underlying physical symmetries are constant, metamorphic relations identified in one ocean-modeling application might be transferred to other applications representing the same physical system. 

In Section~\ref{s-problem}, we formulate the problem to be solved, before introducing the example application in Section~\ref{s-application}.
In Section~\ref{s-relations}, we restrict the problem to finding metamorphic relations that are affine transformations for which we sketch solutions employing machine learning in Section~\ref{s-solution}. 
Section~\ref{s-future} concludes the paper with an outlook to future work.

\section{Formulation of the Problem}\label{s-problem}

Metamorphic testing may be written as
\begin{equation}
    \label{eq:metamorphic_relation}
    f(g(X)) = h(f(X))
\end{equation}
with $X$ denoting all input data to the application under test $f$ and with the pair of functions $(g, h)$ as the metamorphic relation, we seek.
As an initial step, we assume $h$ to be the identity map and we are left with finding $g$, or at least approximations to $g$.
Here we propose to use machine learning algorithms for finding possible $g$.
These algorithms typically minimize a cost function.
Assuming we start from a known set of metamorphic relations $\{g_0, g_1, \ldots, g_{n-1}\}$, we select a cost function that rewards $g_n$ which minimizes $|f(g_n(X)) - f(X)|$ and penalizes if $g_n$ is already known:
\begin{equation}
   \label{eq:cost_function}
   cost(g_n, \{g_0, \ldots, g_{n-1}\}, X) = \frac{|f(g_n(X)) - f(X)|}{\prod_{i=0}^{n-1}\left|g_n(X) - g_i(X)\right|^2}
\end{equation}
We start the iteration with $g_0$ being the identity map and define the norm $\left|X\right|$ to be the square root of the sum of the squares of the atomic data points $X$.

\section{Example Application}\label{s-application}

We provide an example application as an artifact~\cite{Rath2019} that can be directly used on Binder~\cite{Jupyter2018BinderScale.}.
This example shows how to use metamorphic relations derived from the equations describing the applications to uncover bugs in the implementation.

As an example application, we will use the calculation of a time series of the kinetic energy of the surface ocean
\begin{equation}
   \label{eq:energy}
    e(t, y, x) = \frac{1}{2}\frac{\int{\rm d}y{\rm d}x (u(t, y, x)^2 + v(t, y, x)^2)}{\int{\rm d}y{\rm d}x}
\end{equation}
where $t$ indicates time, $y$ and $x$ are spatial coordinates, and  horizontal surface-velocities $u$ and $v$ are calculated from sea-level $ssh$ using
\begin{equation}
    \label{eq:uv}
    (u, v)(t, y, x) = \frac{G}{F} \left(-\frac{\partial}{\partial y}, \frac{\partial}{\partial x}\right) ssh(t, y, x)
\end{equation}
The coordinate system is chosen to have a cyclic boundary condition in both $y$ and $x$ and there are the physical constants $G$ and $F$.
From Equations~\eqref{eq:energy} and \eqref{eq:uv}, it is straightforward to find metamorphic relations based on symmetries of the physical system.
Among these symmetries are sign changes of any of $ssh$, $y$, $x$, $G$, or $F$, scaling of $G$ and $F$ that does not change the ratio $G/F$, or linear transformations of the coordinate system that leave nearest-neighbor relations intact (transposition, translation).

In our example application, there are two implementations of Equation~\eqref{eq:energy} one of which is not respecting cyclic boundary condition.
The two implementations can be found in the artifact~\cite{Rath2019}.

To numerically implement Equation~\eqref{eq:energy}, we discretize the integral in \eqref{eq:energy} and the partial derivatives in \eqref{eq:uv} by applying the method of finite differences.
Linking back to Equation~\eqref{eq:metamorphic_relation}, the input data $X$ contain all discretized values of sea level $ssh$, of all coordinates $t$, $y$, $x$, and the physical constants $G$ and $F$.
A metamorphic transformation $g(x)$ would change all or part of the atomic data points in $X$ such that all $e(g(X)) = e(X)$ for all discrete values of time $t$.

\section{Metamorphic Relations as affine transformations}
\label{s-relations}

We propose to formulate the metamorphic relation $g$ as an affine transformation relating all the atomic data points in $X$ to all the atomic data points in $g(X)$
\begin{equation}
    X' = g(X)
\end{equation}
with
\begin{eqnarray}
    \label{eq:affine_transformation}
    X'_k = \sum_{m} \alpha_{k,m} X_m + \beta_k
\end{eqnarray}
where the $\alpha_{k,m}$ represents a linear mapping from $X$ and $\beta_{k}$ add arbitrary offsets to the atomic data points.
Identifiying metamorphic relations then amounts to finding $\alpha_{k,m}$ and $\beta_{k}$ such that the cost function~\eqref{eq:cost_function} becomes zero for arbitrary input data $X$.

\section{Solution Exploration}\label{s-solution}

Existing machine learning methods are explored to solve Equation~\eqref{eq:affine_transformation} optimized for the cost function~\eqref{eq:cost_function}. 
A modified algorithm loosely based on genetic algorithms with Monte Carlo optimization can be explored.
The initial values for $\{\alpha, \beta\}$ are chosen randomly and in each step a mutation $\delta\{\alpha, \beta\}$ is proposed.
Probability of acceptance of the mutation (p) is specified and iterative mutations applied to solve for the defined cost function \eqref{eq:cost_function}. 
On converging to a local minimum, $\{\alpha, \beta\}$ values are recorded and randomly reset to identify other solutions for equation \eqref{eq:affine_transformation}.

\section{Challenges and Future Work}\label{s-future}

The challenge with these methods would be to explain the metamorphic relations in terms of physical symmetries that hold in the real Ocean.
Nonetheless, we believe that even metamorphic relations which do not reflect symmetries of the physical system can still be used to construct regression tests.

Further work is to relax the restriction to $h$~\eqref{eq:metamorphic_relation} being the identity map and to allow for $g$ not being an affine transformation.
This, however, would generalize $g$ and $h$ to be any non-linear map of the atomic data points which would be challenging to formalize.

\end{document}